\documentclass[onecolumn,showpacs,11pt]{revtex4}
\usepackage{graphicx}
\usepackage{dcolumn}
\usepackage{bm}
\begin{document}
\newcommand{\hs}{\hspace*{0.5cm}}
\newcommand{\vs}{\vspace*{0.5cm}}
\newcommand{\be}{\begin{equation}}
\newcommand{\ee}{\end{equation}}
\newcommand{\bea}{\begin{eqnarray}}
\newcommand{\eea}{\end{eqnarray}}
\newcommand{\ben}{\begin{enumerate}}
\newcommand{\een}{\end{enumerate}}
\newcommand{\bde}{\begin{widetext}}
\newcommand{\ede}{\end{widetext}}
\newcommand{\nn}{\nonumber}
\newcommand{\crn}{\nonumber \\}
\newcommand{\non}{\nonumber}
\newcommand{\noi}{\noindent}
\newcommand{\al}{\alpha}
\newcommand{\la}{\lambda}
\newcommand{\bet}{\beta}
\newcommand{\ga}{\gamma}
\newcommand{\va}{\varphi}
\newcommand{\om}{\omega}
\newcommand{\pa}{\partial}
\newcommand{\fr}{\frac}
\newcommand{\bc}{\begin{center}}
\newcommand{\ec}{\end{center}}
\newcommand{\Ga}{\Gamma}
\newcommand{\de}{\delta}
\newcommand{\De}{\Delta}
\newcommand{\ep}{\epsilon}
\newcommand{\varep}{\varepsilon}
\newcommand{\ka}{\kappa}
\newcommand{\La}{\Lambda}
\newcommand{\si}{\sigma}
\newcommand{\Si}{\Sigma}
\newcommand{\ta}{\tau}
\newcommand{\up}{\upsilon}
\newcommand{\Up}{\Upsilon}
\newcommand{\ze}{\zeta}
\newcommand{\ps}{\psi}
\newcommand{\Ps}{\Psi}
\newcommand{\ph}{\phi}
\newcommand{\vph}{\varphi}
\newcommand{\Ph}{\Phi}
\newcommand{\Om}{\Omega}

\title{The 3-3-1 model with $A_4$ flavor symmetry}

\author{P. V. Dong}
\email {pvdong@iop.vast.ac.vn} \affiliation{Institute of Physics,
VAST, P. O. Box 429, Bo Ho, Hanoi 10000, Vietnam}
\author{L. T. Hue}
\email{lthue@grad.iop.vast.ac.vn} \affiliation{Institute of
Physics, VAST, P. O. Box 429, Bo Ho, Hanoi 10000, Vietnam}
\author{H. N. Long}
\email{hnlong@iop.vast.ac.vn} \affiliation{Institute of Physics,
VAST, P. O. Box 429, Bo Ho, Hanoi 10000, Vietnam}
\author{D. V. Soa}
\email{dvsoa@assoc.iop.vast.ac.vn} \affiliation{Department of
Physics, Hanoi University of Education, Hanoi, Vietnam}

\date{\today}

\begin{abstract}

We argue that the $A_4$ symmetry as required by three flavors of
fermions may well-embed in the $\mathrm{SU}(3)_C\otimes
\mathrm{SU}(3)_L\otimes \mathrm{U}(1)_X$ gauge model. The new
neutral fermion singlets as introduced in a canonical seesaw
mechanism can be combined with the standard model lepton doublets
to perform $\mathrm{SU}(3)_L$ triplets. Various leptoscalar
multiplets such as singlets, doublets, and triplets as played in
the models of $A_4$ are unified in single $\mathrm{SU}(3)_L$
antisextets. As a result, naturally light neutrinos with various
kinds of mass hierarchies are obtained as a combination of type I
and type II seesaw contributions. The observed neutrino mixing
pattern in terms of the Harrison-Perkins-Scott proposal is
obtained by enforcing of the $A_4$ group. The quark masses and
Cabibbo-Kobayashi-Maskawa mixing matrix are also discussed. By
virtue of very heavy antisextets the nature of the vacuum
alignments of scalar fields can be given.

\end{abstract}


\maketitle

\section{\label{intro}Introduction}

The explanation of the smallness of the neutrino masses and the
profile of their mixing as required by experiment
\cite{superK,kam,sno} have been a great puzzle in particle physics
beyond the standard model (SM). The current experimental data are
consistent with the tribimaximal form as proposed by
Harrison-Perkins-Scott \cite{hps}, which, apart from phase
redefinitions, is given by
\begin{eqnarray}
U^{\mathrm{HPS}}=\left(
\begin{array}{ccc}
\frac{2}{\sqrt{6}}       &\frac{1}{\sqrt{3}}  &0\\
-\frac{1}{\sqrt{6}}      &\frac{1}{\sqrt{3}}  &\frac{1}{\sqrt{2}}\\
-\frac{1}{\sqrt{6}}      &\frac{1}{\sqrt{3}}  &-\frac{1}{\sqrt{2}}
\end{array}\right).\label{eq:1}
\end{eqnarray}
It is an interesting challenge to formulate dynamical principles
that can lead to the tribimaximal mixing pattern given in a
completely natural way as a first approximation. Along these lines
the flavor symmetries have been extensively studied. For the first
time, Ma and Rajasekaran \cite{mara} have advocated choosing
$A_4$, the symmetry group of a tetrahedron, as a family symmetry
group. An incomplete list of interesting works that came later
include  Refs. \cite{ma,alta,alta-volkas,A4,yin}. The key to its
success is that the patterns of symmetry breaking with preserved
subgroups are $A_4 \to Z_3$ and $A_4 \to Z_2$ in the two different
sectors- the charged lepton sector and the neutrino sector,
respectively. This misalignment can further be explained by
auxiliary symmetries and particles or even in the context of extra
dimensions (see, for example, \cite{alta-volkas,alta}).

Here we would like to extend the above application to the
$\mathrm{SU}(3)_C\otimes \mathrm{SU}(3)_L \otimes \mathrm{U}(1)_X$
(3-3-1) gauge model \cite{331m,331r,ecn331}, because it can give a
partial explanation of the existence of just three fermion
families in  nature as a result of the gauge anomaly cancellations
required by the $A_4$ symmetry. There are two typical variants of
the 3-3-1 model as far as the lepton sectors are concerned. In the
minimal version, three $\mathrm{SU}(3)_L$ lepton triplets are of
the form $(\nu_L,e_L,e^c_R)_{i=1,2,3}$, where $e_{iR}$ are
ordinary right-handed charged-leptons \cite{331m}. In the second
version, the third components of the lepton triplets include right
-handed neutrinos, respectively, $(\nu_L,e_L,\nu^c_R)_{i=1,2,3}$
\cite{331r}. Note that Ref. \cite{yin} has considered the $A_4$
symmetry in the 3-3-1 model with heavy charged leptons, which is a
modification of the minimal version.

In this work we will pay attention to the second version and try
to recover the tribimaximal form. By analysis, a possibility close
to the typical version is when we replace the right -handed
neutrinos by those with vanishing lepton -number
\cite{matd,mara,malec}. The neutrinos thus gain masses only from
contributions of SU(3)$_L$ scalar antisextets. After considering
the quark sector, the scalar sector is completed. In this model,
the antisextets contain tiny vacuum expectation values (VEVs) in
the first components, as in the case of the standard model with
scalar triplets. To avoid  the decay of the $Z$ boson into the
Majorons associated with these components, the lepton -number
violating scalar-potential should be taken into account.
Therefore, the lepton number is no longer of an exact symmetry;
i.e. the Majorons can get large enough masses to escape from the
decay of $Z$ \cite{matd}. If the antisextets are supposed to be
very heavy, the potential minimization conditions can naturally
give an explanation of the expected vacuum alignments, and also
the smallness of the lepton -number violating VEVs as well as the
mentioned ones. Note that this dangerous decay channel of the $Z$
boson has not been fully  evaluated in the versions of the 3-3-1
model that include the antisextets \cite{tuly-br}.

The paper is organized as follows: In Sec. \ref{model}, we
introduce the $A_4$ family symmetry into the model and obtain the
mass mechanisms and mixing matrix of leptons. Section \ref{quark}
discusses the quark masses. The scalar sector is then completed.
Section \ref{vev} is devoted to the scalar potential, vacuum
alignment problem for the scalar fields. In the last section, Sec.
\ref{conclus}, we summarize our results and make conclusions.
Finally, the appendixes provide the basics of $A_4$ symmetry and
the general scalar potential used in the text.

\section{\label{model}Leptons}

The particle content of the 3-3-1 model under consideration is
collected from  Ref.\cite{mara}. We will show that this selection,
with an appropriate $A_4$ flavor symmetry, provides, in the
framework, a consistent mixing pattern and masses for the
neutrinos. The leptons, under
$(\mathrm{SU}(3)_L,\mathrm{U}(1)_X,A_4)$ symmetries, transform as
\bea \psi_L = \left(%
\begin{array}{c}
  \nu_L \\
  e_L \\
  \nu^c_R \\
\end{array}%
\right)\sim (3,-1/3,\underline{3}),\eea \bea e_{1R}\sim
(1,-1,\underline{1}),\hs e_{2R}\sim (1,-1,\underline{1}'),\hs
e_{3R}\sim (1,-1,\underline{1}''),\eea where $\nu_{iR}$
$(i=1,2,3)$ are three right-handed fermions which are singlets
under the standard model symmetry and have zero lepton number,
$L(\nu_R)=0$. The $X$ charge of the $\mathrm{U}(1)_X$ group is
related to the electric charge operator as
$Q=T_3-\fr{1}{\sqrt{3}}T_8+X$, where $T_a$ $(a=1,2,...,8)$ are
$\mathrm{SU}(3)_L$ charges. Our model is therefore a type of the
ones given in \cite{331m,331r}.

The lepton number in this model does not commute with the gauge
symmetry. It is thus better to work with a new lepton charge
$\mathcal{L}$ related to the lepton number $L$ by diagonal
matrices $L= xT_3+ yT_8 +\mathcal{L}$. Applying $L$ to the lepton
triplet, the coefficients are defined as $x=0$, $y=2/\sqrt{3}$,
and thus $L=\fr{2}{\sqrt{3}}T_8+\mathcal{L}$ \cite{clong}. The
$\mathcal{L}$ charges for the multiplets are as follow:\bc
\begin{tabular}{|c|c|c|c|c|}
  \hline
  Multiplet & $\psi_L$ & $e_{1R}$ & $e_{2R}$ & $e_{3R}$ \\
  \hline
  $\mathcal{L}$ & $2/3$ & 1 & 1 & 1\\
  \hline
\end{tabular}\ec

To generate masses for the charged leptons, we introduce the
following scalar fields: \bea \phi= \left(%
\begin{array}{c}
 \phi^+_1 \\
 \phi^0_2 \\
 \phi^+_3 \\
\end{array}%
\right)\sim (3,2/3,\underline{3},-1/3).\eea The first three
quantum numbers are well -defined as before. The last one is the
$\mathcal{L}$ charge for $\phi$ such that the following Yukawa
interaction is conserved: \bea
\mathcal{L}_l=-h_{ijk}\bar{\psi}_{iL}\phi_j e_{kR}+h.c.,\eea
where \bea h_{ij1}=h_1\left(%
\begin{array}{ccc}
  1 & 0 & 0 \\
  0 & 1 & 0 \\
  0 & 0 & 1 \\
\end{array}%
\right),\hs h_{ij2}=h_2\left(%
\begin{array}{ccc}
  1 & 0 & 0 \\
  0 & \om & 0 \\
  0 & 0 & \om^2 \\
\end{array}%
\right),\hs h_{ij3}=h_3\left(%
\begin{array}{ccc}
  1 & 0 & 0 \\
  0 & \om^2 & 0 \\
  0 & 0 & \om \\
\end{array}%
\right)\eea with $\om= e^{2\pi i/3}$. The lepton number for the
components of $\phi$, including the additional scalars as shown
below, is explicitly given in Appendix \ref{apb}.

The VEV  of $\phi$ is $(v_1,v_2,v_3)$ under $A_4$. The mass
Lagrangian for the charged leptons reads
$\mathcal{L}^{\mathrm{mass}}_l=-(\bar{e}_{1L},\bar{e}_{2L},\bar{e}_{3L})
M_l (e_{1R},e_{2R},e_{3R})^T+h.c.$, where \bea M_l=
\left(%
\begin{array}{ccc}
  h_1v_1 & h_2v_1 & h_3 v_1 \\
   h_1v_2 & h_2\om v_2 & h_3\om^2 v_2 \\
  h_1v_1 & h_2\om^2 v_2 & h_3\om v_3 \\
\end{array}%
\right).\eea We put $v_1=v_2=v_3=v$ so that $A_4$ is broken down
to
$Z_3$ (this is also a minimal condition for the Higgs potential as shown below).
The mass matrix is then diagonalized, \bea U^\dagger_L M_lU_R=\left(%
\begin{array}{ccc}
  \sqrt{3}h_1 v & 0 & 0 \\
  0 & \sqrt{3}h_2 v & 0 \\
  0 & 0 & \sqrt{3}h_3 v \\
\end{array}%
\right)=\left(%
\begin{array}{ccc}
  m_e & 0 & 0 \\
  0 & m_\mu & 0 \\
  0 & 0 & m_\tau \\
\end{array}%
\right),\eea where \bea U_L=\fr{1}{\sqrt{3}}\left(%
\begin{array}{ccc}
  1 & 1 & 1 \\
  1 & \om & \om^2 \\
  1 & \om^2 & \om \\
\end{array}%
\right),\hs U_R=1.\label{lep}\eea

Notice that $\bar{\psi}^c_L \psi_L\phi$ is suppressed because of
the $\mathcal{L}$--symmetry violation. Then $\bar{\psi}^c_L\psi_L$
can couple to SU(3)$_L$ antisextets to generate masses for the
neutrinos. The antisextets in this model transform as \bea \sigma=
\left(%
\begin{array}{ccc}
  \sigma^0_{11} & \sigma^+_{12} & \sigma^0_{13} \\
  \sigma^+_{12} & \sigma^{++}_{22} & \sigma^+_{23} \\
  \sigma^0_{13} & \sigma^+_{23} & \sigma^0_{33} \\
\end{array}%
\right)\sim (6^*,2/3,\underline{1},-4/3), \eea \bea s=
\left(%
\begin{array}{ccc}
  s^0_{11} & s^+_{12} & s^0_{13} \\
  s^+_{12} & s^{++}_{22} & s^+_{23} \\
  s^0_{13} & s^+_{23} & s^0_{33} \\
\end{array}%
\right)\sim (6^*,2/3,\underline{3},-4/3). \eea The Yukawa
interactions are \bea \mathcal{L}_\nu&=&- \fr 1 2
x(\bar{\psi}^c_{1L}\psi_{1L}+\bar{\psi}^c_{2L}\psi_{2L}+\bar{\psi}^c_{3L}\psi_{3L})\sigma\crn
&& -
y(\bar{\psi}^c_{2L}\psi_{3L}s_1+\bar{\psi}^c_{3L}\psi_{1L}s_2+\bar{\psi}^c_{1L}\psi_{2L}s_3)\crn
&&+h.c.\eea

The VEV of $s$ is set as $(\langle s_1\rangle,0,0)$ under $A_4$
(which is also a natural minimal condition for the Higgs
potential). As such, the group is broken down to $Z_2$ in the
neutrino sector, where \bea
\langle s_1\rangle=\left(%
\begin{array}{ccc}
  u'_{1} & 0 & u_{1} \\
  0 & 0 & 0 \\
  u_{1} & 0 & \Lambda_{1} \\
\end{array}%
\right).\eea The VEV of $\sigma$ is \bea
\langle \sigma \rangle=\left(%
\begin{array}{ccc}
  u' & 0 & u \\
  0 & 0 & 0 \\
  u & 0 & \La \\
\end{array}%
\right).\eea The mass Lagrangian for the neutrinos is defined by
\bea \mathcal{L}^{\mathrm{mass}}_\nu=-\fr 1 2 \bar{\chi}^c_L M_\nu
\chi_L+ h.c.,\hs  \chi_L\equiv
\left(%
\begin{array}{c}
  \nu_L \\
  \nu^c_R \\
\end{array}%
\right),\hs M_\nu\equiv\left(%
\begin{array}{cc}
  M_L & M^T_D \\
  M_D & M_R \\
\end{array}%
\right),\label{nm}\eea where $\nu=(\nu_{1},\nu_{2},\nu_{3})^T$.
The mass matrices are then obtained by
\bea M_{L,R,D}=\left(%
\begin{array}{ccc}
  a_{L,R,D} & 0 & 0 \\
  0 & a_{L,R,D} & b_{L,R,D} \\
  0 & b_{L,R,D} & a_{L,R,D} \\
\end{array}%
\right),\eea with \bea a_L=x u',\ a_D=x u,\ a_R=x \La,\ b_L=y
u'_{1},\ b_D= y u_1,\ b_R=y \La_1.\eea

Three active -neutrinos gain masses via a combination of type I
and type II seesaw mechanisms derived from (\ref{nm}) as \bea
M^{\mathrm{eff}}=M_L-M_D^TM_R^{-1}M_D=
\left(%
\begin{array}{ccc}
  a' & 0 & 0 \\
  0 & a & b \\
  0 & b & a \\
\end{array}%
\right),\label{neu}\eea where \bea a'&=&a_L-\fr{a^2_D}{a_R},\crn
a&=&a_L+2a_Db_D\fr{b_R}{a_R^2-b^2_R}-(a^2_D+b^2_D)\fr{a_R}{a_R^2-b^2_R},\crn
b&=&b_L-2a_Db_D\fr{a_R}{a_R^2-b^2_R}+(a^2_D+b^2_D)\fr{b_R}{a_R^2-b^2_R}.\label{neu3}\eea

We can diagonalize the mass matrix (\ref{neu}) as follows: \bea
U^T_\nu M^{\mathrm{eff}} U_\nu=\left(%
\begin{array}{ccc}
  a+b & 0 & 0 \\
  0 & a' & 0 \\
  0 & 0 & a-b \\
\end{array}%
\right)=\left(%
\begin{array}{ccc}
  m_1 & 0 & 0 \\
  0 & m_2 & 0 \\
  0 & 0 & m_3 \\
\end{array}%
\right),\label{neu2} \eea where \bea U_\nu=\left(%
\begin{array}{ccc}
  0 & 1 & 0 \\
  \fr{1}{\sqrt{2}} & 0 & -\fr{1}{\sqrt{2}} \\
  \fr{1}{\sqrt{2}} & 0 & \fr{1}{\sqrt{2}} \\
\end{array}%
\right).\label{neu1}\eea Combined with (\ref{lep}), the lepton
mixing matrix yields the tribimaximal mixing pattern as proposed
by Harrison- Perkins- Scott (up to a phase): \bea U^\dagger_L
U_\nu=\left(%
\begin{array}{ccc}
  \sqrt{2/3} & 1/\sqrt{3} & 0 \\
  -1/\sqrt{6} & 1/\sqrt{3} & i/\sqrt{2} \\
  -1/\sqrt{6} & 1/\sqrt{3} & -i/\sqrt{2} \\
\end{array}%
\right)=U^{\mathrm{HPS}}P_\phi,\eea where the phase matrix
$P_\phi=\mathrm{diag}(1,1,i)$ can be removed by absorbing it into
the neutrino mass eigenstates. This is a main result of the paper.

With the aid of the results in (\ref{neu3}), we identify $u',u'_1$
as the VEVs of the type II seesaw mechanism. The mechanism works
because, from Eq. (\ref{scales}) in Sec. \ref{vev}, the
spontaneous breaking of electroweak symmetry is already
accomplished by $v$; hence $u',u'_1$ may be small, as long as $M$
is large. The parameter $\bar{\mu}_2$ (which has the dimension of
mass) may also be naturally small, because its absence enhances
the symmetry of $V^{s\sigma}$ \cite{matd}. On the other hand,
$u,u_1$ are the VEV of the type I seesaw mechanism. Similar to the
case above, these VEVs are, however, much smaller than $v$. But
they can be larger than $u',u'_1$ because $v_\chi
> v$ (notice that $v_\chi$ is the scale of the 3-3-1 symmetry breaking into the SM). The
TeV scale type I seesaw mechanism can be achieved if we take
$v_\chi=10$ TeV, $\bar{\mu}_1 = 100 \bar{\mu}_2$ \cite{matd}.

It is noted that the lepton number $L$ is really broken by the
small VEVs of the antisextets $s_1$ and $\sigma$ since their
corresponding field components carry $L$; namely the (11) has $L =
-2$, the (13) has $L = -1$, but the (33) has $L=0$. Now $u'\neq 0$
(or $u'_1 \neq 0$) by itself means that $L$ is broken by 2 units;
hence $L\rightarrow (-)^L$, as lepton parity is still conserved.
This is the case in most models of neutrino mass. The type I
seesaw mechanism gives no contribution. However, if $u$ (or $u_1$)
is also nonzero, then $L$ is broken completely. Both the seesaw
mechanisms play this role.

\section{\label{quark}Quarks}

It is well known that the 3-3-1 model is a good example  of the
fermion number problem: Why are there only three families of
fermions in nature \cite{331m,331r,clong}? This perfectly meets
the criteria of three-family symmetry theories such as $A_4$. The
anomaly cancellation in the 3-3-1 models requires the number of
$\mathrm{SU}(3)_L$ triplets to be equal to the number of
$\mathrm{SU}(3)_L$ antitriplets; i.e.,  two families of quarks
have to transform differently from the other one. Hence, the quark
triplets and antitriplets of the three families cannot lie in a
\underline{3} representation of $A_4$. The right-handed exotic
quarks are the same. Here, the following two situations exist .

The first situation is that the above scalar $\phi$ is
\emph{responsible} for generating quark masses. The quark content
is obtained as follows: \bea Q_{3L}&=&
\left(%
\begin{array}{c}
  u_{3L} \\
  d_{3L} \\
  T_L \\
\end{array}%
\right)\sim (3,1/3,\underline{1},-1/3),\label{pt1}\\ Q_{1L}&=&
\left(%
\begin{array}{c}
  d_{1L} \\
  -u_{1L} \\
  D_{1L} \\
\end{array}%
\right)\sim (3^*,0,\underline{1}',1/3),\hs Q_{2L}=
\left(%
\begin{array}{c}
  d_{2L} \\
  -u_{2L} \\
  D_{2L} \\
\end{array}%
\right)\sim (3^*,0,\underline{1}'',1/3),\label{pt3}\\ T_R&\sim&
(1,2/3,\underline{1},-1),\hs D_{1R}\sim
(1,-1/3,\underline{1}'',1),\hs D_{2R}\sim
(1,-1/3,\underline{1}',1),\label{pt4}\\ u_R &\sim&
(1,2/3,\underline{3},0),\hs d_R \sim
(1,-1/3,\underline{3},0).\label{pt2}\eea From (\ref{pt1}),
(\ref{pt3}) and (\ref{pt4}), it follows that the exotic quarks
have single lepton number, i.e. $L(T)= -1$ and $L(D)= +1$. Hence,
in the considered model the exotic quarks are leptoquarks. With
the above quark content, the scalar triplet $\phi$ is not enough
to provide mass for all the quarks. Hence, the following extra
scalar fields are needed to provide masses for the remaining
quarks~\cite{331r}: \bea \eta=
\left(%
\begin{array}{c}
  \eta^0_1 \\
  \eta^-_2 \\
  \eta^0_3 \\
\end{array}%
\right)\sim (3,-1/3,\underline{3},-1/3),\hs \chi=\left(%
\begin{array}{c}
  \chi^0_1 \\
  \chi^-_2 \\
  \chi^0_3 \\
\end{array}%
\right)\sim (3,-1/3,\underline{1},2/3).\eea The Yukawa
interactions are \bea -\mathcal{L}_q &=& h^d_3 \bar{Q}_{3L}(\phi
d_R)_1 + h^u_1 \bar{Q}_{1L}(\phi^*u_R)_{1''}+ h^u_2
\bar{Q}_{2L}(\phi^*u_R)_{1'}\crn &&+ h^u_3 \bar{Q}_{3L}(\eta
u_R)_1+h^d_1 \bar{Q}_{1L}(\eta^* d_R)_{1''}+h^d_2
\bar{Q}_{2L}(\eta^* d_R)_{1'}\crn && + f_3 \bar{Q}_{3L}\chi T_R +
f_1 \bar{Q}_{1L}\chi^* D_{1R}+f_2 \bar{Q}_{2L}\chi^* D_{2R}\crn
&&+h.c.\eea

Suppose that the VEVs of $\eta$ and $\chi$ are $(v',v',v')$ and
$v_\chi$, with  $v'=\langle \eta^0_1\rangle$, $v_\chi=\langle
\chi^0_3\rangle$, $\langle \eta^0_3\rangle=0$, and
$\langle\chi^0_1\rangle=0$. The exotic quarks get masses directly
from the VEV of $\chi$: $m_T=f_3 v_\chi$, $m_{D_{1,2}}=f_{1,2}
v_\chi$. In addition, $v_\chi$ has to be much larger than those of
$\phi$ and $\eta$. The mass matrices for ordinary up -quarks and
down -quarks are, respectively, obtained as follows: \bea M_u =
\left(%
\begin{array}{ccc}
  -h^u_1 v & -h^u_1 \om v  & -h^u_1 \om^2 v  \\
   -h^u_2 v & -h^u_2 \om^2 v  & -h^u_2 \om v   \\
  h^u_3 v' & h^u_3 v' & h^u_3 v' \\
\end{array}%
\right),\hs M_d=
\left(%
\begin{array}{ccc}
  h^d_1 v' & h^d_1 \om v'  & h^d_1 \om^2 v'  \\
   h^d_2 v' & h^d_2 \om^2 v'  & h^d_2 \om v'   \\
  h^d_3 v & h^d_3 v & h^d_3 v \\
\end{array}%
\right). \eea Let us put \bea A=
\fr{1}{\sqrt{3}}\left(%
\begin{array}{ccc}
  1 & 1 & 1 \\
  \om^2 & \om & 1 \\
  \om & \om^2 & 1 \\
\end{array}%
\right).\eea We have then \bea M_u A&=&
\left(%
\begin{array}{ccc}
  -\sqrt{3}h^u_1 v & 0 & 0 \\
  0 & -\sqrt{3}h^u_2 v & 0 \\
  0 & 0 & \sqrt{3}h^u_3 v' \\
\end{array}%
\right)=\left(%
\begin{array}{ccc}
  m_u & 0 & 0 \\
  0 & m_c & 0 \\
  0 & 0 & m_t \\
\end{array}%
\right), \crn M_d A&=&
\left(%
\begin{array}{ccc}
  \sqrt{3}h^d_1 v' & 0 & 0 \\
  0 & \sqrt{3}h^d_2 v' & 0 \\
  0 & 0 & \sqrt{3}h^d_3 v \\
\end{array}%
\right)
=\left(%
\begin{array}{ccc}
  m_d & 0 & 0 \\
  0 & m_s & 0 \\
  0 & 0 & m_b \\
\end{array}%
\right).\eea The unitary matrices, which couple the left-handed
up- and down -quarks to those in the mass bases, are $U^u_L=1$ and
$U^d_L=1$, respectively. Therefore we get the
Cabibbo-Kobayashi-Maskawa (CKM) matrix \bea
U_\mathrm{CKM}=U^{d\dagger}_L U^u_L=1.\label{a41}\eea Note that
the property in (\ref{a41}) is common for some models based on the
$A_4$ group.

In the last situation: the mentioned scalar field $\phi$  \emph{is
not responsible} for the quark masses. The ordinary right-handed
quarks are therefore in singlets under $A_4$. In this case, we
might introduce three extra $\mathrm{SU}(3)_L$ Higgs triplets such
as \bea \eta&=&
          \left(
            \begin{array}{c}
              \eta^0_1 \\
              \eta^-_2 \\
              \eta^0_3 \\
            \end{array}
          \right)\sim (3,-1/3,\underline{1},-1/3),\hs
          \rho=\left(
            \begin{array}{c}
              \rho^+_1 \\
              \rho^0_2 \\
              \rho^+_3 \\
            \end{array}
          \right)\sim (3,2/3,\underline{1},-1/3),
          \eea and $\chi$, as in  the first situation .
A combination of such Higgs scalar fields will give mass for  all
the quarks~\cite{331r}. However, all these scalar triplets as well
as the quarks lie in \underline{1} representations of $A_4$. It is
easy to check that all quarks get masses in the same ordinary
3-3-1 model; namely, $v_\eta=\langle \eta^0_1\rangle$ provides the
mass for $u_3$, $d_1$, and $d_2$ quarks,
 $v_\rho=\langle \rho^0_2\rangle$ for $d_3$, $u_1$, and $u_2$
quarks, and $v_\chi=\langle\chi^0_3\rangle$ for exotic quarks $T$,
$D_1$, and $D_2$.

Notice that, for both situations, if the lepton parity $(-)^L$ is
broken, i.e. the lepton number $L$ is broken completely, then
there is no longer a symmetry which protects $\eta^0_3$ $(L=-1)$
and $\chi_1^0$ $(L=1)$ from acquiring VEVs. This will induce
mixing between the leptoquarks and the usual quarks, which may
lead to the effects of flavor changing neutral currents. This kind
of mixing in the 3-3-1 model has been studied in a number of
papers \cite{qmixng}, so we will not discuss it further. Anyway,
the solution corresponding to the residual symmetry $(-)^L$ should
be more natural.

In this model the first situation is quite natural because the
$A_4$ triplet $\eta$, which may strongly couple to $\phi$ via some
potential, will be aligned in the $(1,1,1)$ VEV direction of
$\phi$, as assumed. Namely, we can check that those VEV structures
for $\phi$ and $\eta$ are an automatic solution from the potential
minimization conditions; no misalignment solution appears. But, in
the following we will consider the scalar and quark content of the
second situation. The results obtained can be similarly derived
for the first situation. The scalar content and general scalar
potential in the case of interest are summarized in
appendix~\ref{apb}. Note that, in Ref. \cite{yin}, only the lepton
sector has been considered, and the quark sector has not been
mentioned.

\section{\label{vev}Vacuum Alignment}

There are several scalar sectors where $\phi$ is responsible for
charged lepton masses, $\sigma$ and $s$ are responsible for
neutrino masses, and $\eta$, $\rho$, $\chi$ -for quark masses,
with the vacuum structures  shown above. If the first two sectors
such as $\phi$ and $s$ are strongly coupled, i.e. the couplings of
$V(s,\phi)$ in (\ref{vsp}) are turned on with enough strength,
such vacuum alignments for $\phi$ and $s$ would be broken. To
resolve this problem, we might include extra dimensions as in
\cite{alta} or supersymmetry as in \cite{alta-volkas}. However, in
this work we will provide an alternative explanation, following
\cite{mara,ma}.

At the low-energy limit, the antisextets $\sigma$ and $s$ are
decomposed into the ones of standard model symmetry. Noting that
$6^*=3^*\oplus 2^* \oplus 1$
under $\mathrm{SU}(2)_L$ we get \bea \sigma = \left(%
\begin{array}{cc}
  \sigma^0_{11} & \sigma^+_{12} \\
  \sigma^+_{12} & \sigma^{++}_{22} \\
\end{array}%
\right)\oplus \left(%
\begin{array}{c}
  \sigma^0_{13} \\
  \sigma^+_{23} \\
\end{array}%
\right)\oplus \sigma^0_{33},\hs s=\left(%
\begin{array}{cc}
  s^0_{11} & s^+_{12} \\
  s^+_{12} & s^{++}_{22} \\
\end{array}%
\right)\oplus \left(%
\begin{array}{c}
  s^0_{13} \\
  s^+_{23} \\
\end{array}%
\right)\oplus s^0_{33},\eea where the antitriplets have the lepton
number $L=-2$, antidoublets $L=-1$, and singlets $L=0$. Our
effective theory thus plays the same role as the previously
well-known proposals of $A_4$ such as in Refs. \cite{mara,ma}. The
dynamics of the antitriplets and antidoublets can further  be
found
 in \cite{matd}. Similar to those cases, $\sigma$ and
$s$ in the model maybe very heavy which are all integrated away,
so they do not appear as physical particles at or below the TeV
scale. They have interactions among themselves similar to those of
the potentials for $\phi$ as shown below. Only their imprint at
the low energy is the VEV structures as given.

To see this, let us suppose that the antisextets $\sigma$ and $s$
are heavy,  with masses $\mu_\sigma$ and $\mu_s$, respectively,
and consider the minimization conditions of a potential
$V^{s\sigma}$ concerning to these antisextets. To obtain the
desirable solution $\langle\sigma\rangle \neq 0$, $\langle
s_1\rangle \neq 0$, and $\langle s_2\rangle = \langle s_3\rangle
=0$, the lepton number $\mathcal{L}$, as well as $A_4$, must be
broken as given in (\ref{vi}). The new observation is that the
following choice of soft scalar terms of (\ref{vi}) works in the
$V^{s\sigma}$ potential:\bea
V^{s\sigma}&=&V(s)+V(\sigma)+V(s,\sigma)\crn
&&+\left(\bar{\mu}_{1} \eta^T \sigma \chi+\bar{\mu}_{2} \eta^T
 \sigma\eta+\bar{\la}_{1} \eta^\dagger
 s^\dagger_1\chi\rho+\bar{\la}_{2} \eta^\dagger
 s^\dagger_1\eta\rho+\bar{\la}_{3} \chi^\dagger
 s^\dagger_1\chi\rho+h.c.\right)\eea
 From $V^{s\sigma}$, one solution to
 the minimization conditions is $\langle s_2\rangle =\langle s_3\rangle =0$, and
\bea
\langle s_1\rangle =\left(%
\begin{array}{ccc}
  u'_1 & 0 & u_1 \\
  0 & 0 & 0 \\
  u_1 & 0 & \La_1 \\
\end{array}%
\right),\hs \langle \sigma\rangle =\left(%
\begin{array}{ccc}
  u' & 0 & u \\
  0 & 0 & 0 \\
  u & 0 & \La \\
\end{array}%
\right).\eea Here $\langle s_1\rangle $ and $\langle \sigma\rangle
$ are the root of the $\pa V^{s\sigma}_{\mathrm{min}}/\pa \langle
s_1\rangle ^*=0$ and $\pa V^{s\sigma}_{\mathrm{min}}/\pa \langle
\sigma\rangle ^*=0$ (with $V^{s\sigma}_{\mathrm{min}}$ the minimum
of $V^{s\sigma}$), whereas other similar conditions vanish due to
$\langle s_2\rangle =\langle s_3\rangle =0$. This is also an
important result of our paper.

Since $\La$, $\La_1$ are much larger than $u,u',u_1,u'_1$, from the
minimization conditions $\pa V^{s\sigma}_{\mathrm{min}}/\pa
\La_1^*=0$ and $\pa V^{s\sigma}_{\mathrm{min}}/\pa \La^*=0$ we
derive: \bea
\La^2_1&\simeq&\left[2(\la^\sigma+\la'^s)\mu^2_s-(2\la'^{s\sigma}_3+2\la^{s\sigma}_3
+\la'^{s\sigma}_1+\la^{s\sigma}_1+\la^{s\sigma}_2+\la'^{s\sigma}_2)\mu^2_\sigma\right]
/\left[(2\la'^{s\sigma}_3+2\la^{s\sigma}_3\right.\crn
&&\left.+\la'^{s\sigma}_1+\la^{s\sigma}_1+\la^{s\sigma}_2+\la'^{s\sigma}_2)(\la'^{s\sigma}_2
+\la'^{s\sigma}_1+\la^{s\sigma}_2+\la^{s\sigma}_1+2\la'^{s\sigma}_3+2\la^{s\sigma}_3)
-4(\la^\sigma+\la'^s)\right.\crn
&&\left.\times(\la'^s_1+\la'^s_2+\la^s_1+\la^s_2)\right],\\
\La^2&\simeq&\left[2(\la'^s_1+\la'^s_2+\la^s_1+\la^s_2)\mu^2_\sigma-(\la'^{s\sigma}_2+
\la'^{s\sigma}_1+\la^{s\sigma}_2+\la^{s\sigma}_1+2\la'^{s\sigma}_3+2\la^{s\sigma}_3)\mu^2_s\right]
/\left[(2\la'^{s\sigma}_3\right.\crn
&&\left.+2\la^{s\sigma}_3+\la'^{s\sigma}_1+\la^{s\sigma}_1+\la^{s\sigma}_2+\la'^{s\sigma}_2)(\la'^{s\sigma}_2
+\la'^{s\sigma}_1+\la^{s\sigma}_2+\la^{s\sigma}_1+2\la'^{s\sigma}_3+2\la^{s\sigma}_3)
\right.\crn
&&\left.-4(\la^\sigma+\la'^s)(\la'^s_1+\la'^s_2+\la^s_1+\la^s_2)\right],\eea
i.e. $\La$ and $\La_1$ are on the scale of the antisextets  masses
$\mu_{\sigma}$, $\mu_s$. However,  $u$, $u_1$, $u'$, and $u'_1$
get very small values \cite{matd,ma} derived from the remaining
minimization conditions as given by \bea u' &\simeq&
\fr{[\mu_s^2+2(\la'^s_1 +\la'^s_2)\La^2_1 +\la'^{s\sigma}_1
\La^2]v^2_\eta\bar{\mu}_2+(\la'^{s\sigma}_2+2\la'^{s\sigma}_3)\La\La_1v_\eta
v_\rho v_\chi  \bar{\la}_1}{(\mu^2_\sigma+2\la'^\sigma
\La^2+\la'^{s\sigma}_1 \La^2_1)[\mu_s^2+2(\la'^s_1
+\la'^s_2)\La^2_1 +\la'^{s\sigma}_1
\La^2]-(\la'^{s\sigma}_2+2\la'^{s\sigma}_3)^2\La^2\La^2_1},\label{scales1}\\
u'_1 &\simeq& \fr{-(\la'^{s\sigma}_2+2\la'^{s\sigma}_3)\La\La_1
v_\eta^2 \bar{\mu}_2-(\mu^2_\sigma+2\la'^\sigma
\La^2+\la'^{s\sigma}_1 \La^2_1)v_\eta v_\rho v_\chi
\bar{\la}_1}{(\mu^2_\sigma+2\la'^\sigma \La^2+\la'^{s\sigma}_1
\La^2_1)[\mu_s^2+2(\la'^s_1 +\la'^s_2)\La^2_1 +\la'^{s\sigma}_1
\La^2]-(\la'^{s\sigma}_2+2\la'^{s\sigma}_3)^2\La^2\La^2_1},\label{scales2}\\
u&\simeq&\left\{[2\mu_s^2+4(\la^s_1+\la'^s_1+\la^s_2+\la'^s_2)\La_1^2+(\la^{s\sigma}_1
+2\la'^{s\sigma}_1+\la^{s\sigma}_2)\La^2]v_\eta v_\chi
\bar{\mu}_1\right.\crn && \left.-(\la^{s\sigma}_1+\la^{s\sigma}_2
+2\la'^{s\sigma}_2+4\la^{s\sigma}_3+4\la'^{s\sigma}_3)\La\La_1
v_\rho(v_\eta^2 \bar{\la}_2-v_\chi^2
\bar{\la}_3)\right\}/\left\{[2\mu^2_\sigma+4(\la^\sigma+\la'^{\sigma})\La^2\right.\crn
&& \left.+(\la_1^{s\sigma}
+2\la'^{s\sigma}_1+\la_2^{s\sigma})\La^2_1]
[2\mu_s^2+4(\la^s_1+\la'^s_1+\la^s_2+\la'^s_2)\La_1^2+(\la^{s\sigma}_1+2\la'^{s\sigma}_1+\la^{s\sigma}_2)\La^2]
\right.\crn &&\left.-(\la^{s\sigma}_1+\la^{s\sigma}_2
+2\la'^{s\sigma}_2+4\la^{s\sigma}_3+4\la'^{s\sigma}_3)^2\La^2\La^2_1\right\},\label{scales3}\\
u_1&\simeq&\left\{[2\mu^2_\sigma+4(\la^\sigma+\la'^{\sigma})\La^2+(\la_1^{s\sigma}
+2\la'^{s\sigma}_1+\la_2^{s\sigma})\La^2_1]v_\rho(v_\eta^2
\bar{\la}_2-v_\chi^2 \bar{\la}_3)\right.\crn &&
\left.-(\la^{s\sigma}_1+\la^{s\sigma}_2
+2\la'^{s\sigma}_2+4\la^{s\sigma}_3+4\la'^{s\sigma}_3)\La\La_1v_\eta
v_\chi
\bar{\mu}_1\right\}/\left\{[2\mu^2_\sigma+4(\la^\sigma+\la'^{\sigma})\La^2\right.\crn
&& \left.+(\la_1^{s\sigma}
+2\la'^{s\sigma}_1+\la_2^{s\sigma})\La^2_1]
[2\mu_s^2+4(\la^s_1+\la'^s_1+\la^s_2+\la'^s_2)\La_1^2+(\la^{s\sigma}_1+2\la'^{s\sigma}_1+\la^{s\sigma}_2)\La^2]
\right.\crn &&\left.-(\la^{s\sigma}_1+\la^{s\sigma}_2
+2\la'^{s\sigma}_2+4\la^{s\sigma}_3+4\la'^{s\sigma}_3)^2\La^2\La^2_1\right\}.\label{scales4}
\eea Let us put $\La,\La_1,\mu_\sigma,\mu_s\sim M$. Suppose that
all the terms existing in the same numerator are the same order,
i.e. $v_\eta\sim v_\rho(\sim v)$, $v_\chi\bar{\la}_1\sim
\bar{\mu}_2$, and $v_\chi\bar{\la}_3\sim \bar{\mu}_1$. We derive
\be  u'\sim u'_1\sim \fr{\bar{\mu}_2v^2}{M^2},\hs u\sim u_1\sim
\fr{\bar{\mu}_1vv_\chi}{M^2}.\label{scales} \ee (See also the
remarks in Sec. \ref{model} for completion.)

The potential concerning  $\phi$, after integrating out over the
heavy fields as mentioned, can be identified as
$V^\phi=V(\phi)+V(\phi,\rho)+V(\phi,\eta)+V(\phi,\chi)$. The
minimum of the potential is given by \bea
V^\phi_{\mathrm{min}}&=&(m^2-2\la^{\phi\rho}_3v^2_\rho)(|v_1|^2+|v_2|^2+|v_3|^2)+\la^\phi_1(|v_1|^2+|v_2|^2+|v_3|^2)^2\crn
&& +\la^\phi_2(|v_1|^2+\om^2 |v_2|^2+\om |v_3|^2)(|v_1|^2+\om
|v_2|^2+\om^2 |v_3|^2)\crn && + \la_3^\phi
(|v_2|^2|v_3|^2+|v_3|^2|v_1|^2+|v_1|^2|v_2|^2)+\left\{\la^\phi_4(v^{*}_2v_3)^2+(v^{*}_3v_1)^2+(v^{*}_1v_2)^2
\right.\crn &&
+\left.\la^{\phi\rho}_3v^2_\rho[(v^*_1)^2+(v^*_2)^2+(v^*_3)^2]
+(\la^{\phi\rho}_4+\la^{\phi\rho}_5)v^*_\rho[v^*_1v_2v_3+v_1v^*_2v_3+v_1v_2v^*_3]\right.\crn
&&\left.+c.c.\right\}\eea Here we have defined
$m^2=\mu_\phi^2+\la^{\phi\eta}_1 |v_\eta|^2+\la^{\phi\chi}_1
|v_\chi|^2+(\la^{\phi\rho}_1+\la^{\phi\rho}_2)
|v_\rho|^2+2\la^{\phi\rho}_3v^2_\rho$, with
$v_\eta=\langle\eta\rangle$, $v_\rho=\langle\rho\rangle$ and
$v_\chi=\langle\chi\rangle$. The minimization conditions on $v_i$
are given by \bea \fr{\pa V^\phi_{\mathrm{min}}}{\pa
v^*_1}&=&(m^2-2\la^{\phi\rho}_3v^2_\rho) v_1+2\la^\phi_1
v_1(|v_1|^2+|v_2|^2+|v_3|^2) +\la^\phi_2
v_1(2|v_1|^2-|v_2|^2-|v_3|^2)\crn && +\la^\phi_3 v_1
(|v_2|^2+|v_3|^2) +2\la^\phi_4
v^*_1(v_2^2+v_3^2)+2\la^{\phi\rho}_3v^2_\rho v^*_1 \crn
&&+(\la^{\phi\rho}_4+\la^{\phi\rho}_5)[v^*_\rho v_2v_3 +v_\rho
(v_2^* v_3+v_2 v^*_3)], \eea and other similar equations. One
solution to these equations is \be
v_1=v_2=v_3=\fr{-3v_\rho(\la^{\phi\rho}_4+\la^{\phi\rho}_5)+\sqrt{9|v_\rho|^2(\la^{\phi\rho}_4
+\la^{\phi\rho}_5)^2-8m^2(3\la^\phi_1+\la^\phi_3+2\la^\phi_4)}}{4(3\la^\phi_1+\la^\phi_3+2\la^\phi_4)}.\ee
Let us note that such vacuum alignment does not change when the
terms $\phi$ in (\ref{vi}), except for those coupled to $s$, are
included.

\section{\label{conclus}Conclusions}

We have constructed the $\mathrm{SU}(3)_C\otimes
\mathrm{SU}(3)_L\otimes \mathrm{U}(1)_X$ gauge model based on
$A_4$ flavor symmetry. This 3-3-1 model is different from previous
proposals \cite{331m,331r,yin} because it includes the new neutral
fermion singlets with zero lepton-number following \cite{mara}
into the third components of the $\mathrm{SU}(3)_L$ lepton
triplets, as well as the scalar antisextets as required to
generate the masses for the neutrinos.

The charged leptons gain masses from the Yukawa interactions of
the SU(3)$_L$ triplet $\phi$. The neutrinos and neutral fermion
singlets gain masses from contributions of the antisextets
$\sigma$ and $s$. The three active neutrinos have naturally small
masses as a result of interplay of type I and II seesaw
mechanisms. The quark masses exist in one of the two cases. The
first case is induced by contributions from $\phi$, where the CKM
matrix may be unity at the first approximation. In contrast, the
second case is due to a discriminative scalar sector of the
$\eta,\rho,\chi$ triplets. The resulting masses and mixing matrix
of quarks are  the same as the ordinary 3-3-1 model.

The separation of the two $A_4$ triplets $\phi$ and $s$, which
generate masses for charged leptons and neutrinos respectively,
are evaluated. We have shown that if the antitriplets $\sigma$ and
$s$ are heavy, lepton-number violating vacuum expectation values
maybe induced via the lepton number violating scalar potentials as
well as the scalar soft -terms of $A_4$. The vacuum alignment for
these antisextets exists as a result. The scalar potential
concerning $\phi$ at or below the TeV scale is obtained by
integrating out from the very heavy antisextets, which naturally
yields the vacuum structures as expected. Remember that in this
case the type I seesaw scale is very large, corresponding to those
of the antisextets. To achieve  a TeV seesaw scale, other
mechanisms, such as ones \cite{alta,alta-volkas} for separating
$\phi$ and $s$, should be used.

Finally, since in our model one family of quarks is different from
the  other two, other flavor symmetry groups which contain
\underline{2}-representations such as $S_4$ may be preferred. This
subject is dedicated to future studies.

 \section*{Acknowledgments}
This work was supported in part by the National Foundation for
Science and Technology Development (NAFOSTED) of Vietnam under
Grant No. 103.01.15.09.
\\[0.3cm]

\appendix

\section{\label{apa}$\emph{A}_4$ Symmetry}

For three families of fermions, we should look for a group with an
irreducible \underline{3} representation which acts on the family
indices, the simplest of which is $A_4$, the group of even
permutation of four objects. It is also the symmetry group of a
regular tetrahedron.

The group has 12 elements and four equivalence classes with three
inequivalent one-dimensional represenations and one
three-dimensional one. Its character table is given in Table
\ref{a4}.
\begin{table}[h]
\begin{center}
\begin{tabular}{|c|c|c|c|c|c|}
\hline class & $n$ & $\chi_1$ & $\chi_{1'}$ & $\chi_{1''}$ & $\chi_3$ \\
\hline
 $C_1$ & 1  & 1 & 1 & 1 & 3 \\
 $C_2$ & 4  & 1 & $\omega$ & $\omega^2$ & 0 \\
 $C_3$ & 4  & 1 & $\omega^2$ & $\omega$ & 0 \\
 $C_4$ & 3 & 1 & 1 & 1 & $-1$ \\
\hline
\end{tabular}
\caption{\label{a4} Character table of A$_4$, where
$\omega=e^{2\pi i/3}$ is the cube root of unity.}
\end{center}
\end{table} The multiplication rule for \underline{3} representations is
\begin{eqnarray}
\underline{3} \otimes \underline{3} &=& \underline{1} (11+22+33)
\oplus \underline{1}' (11 + \omega^2 22 + \omega 33) \oplus
\underline{1}'' (11 + \omega 22 + \omega^2 33) \nonumber \\
&&\oplus \underline{3} (23, 31, 12) \oplus \underline{3} (32, 13,
21).
\end{eqnarray} Further, we can denote, on the right - hand
side, the first \underline{3} as $\underline{3_s}$ and the second
\underline{3} as $\underline{3_a}$.

\section{\label{apb}Scalar sector}

\subsection{Scalar content}

Let us summarize the Higgs content of the model: \bea
\phi&=&\left(
            \begin{array}{c}
              \phi^+_1 \\
              \phi^0_2 \\
              \phi^+_3 \\
            \end{array}
          \right)\sim (3,2/3,\underline{3},-1/3),\\ \eta&=&
          \left(
            \begin{array}{c}
              \eta^0_1 \\
              \eta^-_2 \\
              \eta^0_3 \\
            \end{array}
          \right)\sim (3,-1/3,\underline{1},-1/3),\\
          \rho&=&\left(
            \begin{array}{c}
              \rho^+_1 \\
              \rho^0_2 \\
              \rho^+_3 \\
            \end{array}
          \right)\sim (3,2/3,\underline{1},-1/3),\\
          \chi&=&\left(
            \begin{array}{c}
              \chi^0_1 \\
              \chi^-_2 \\
              \chi^0_3 \\
            \end{array}
          \right)\sim (3,-1/3,\underline{1},2/3),\\
          \sigma&=&\left(
                     \begin{array}{ccc}
                       \sigma^0_{11} & \sigma^+_{12} & \sigma^0_{13} \\
                       \sigma^+_{12} & \sigma^{++}_{22} & \sigma^+_{23} \\
                       \sigma^0_{13} & \sigma^+_{23} & \sigma^0_{33} \\
                     \end{array}
                   \right)\sim (6^*,2/3,\underline{1},-4/3),\\
                   s&=&\left(
                     \begin{array}{ccc}
                       s^0_{11} & s^+_{12} & s^0_{13} \\
                       s^+_{12} & s^{++}_{22} & s^+_{23} \\
                       s^0_{13} & s^+_{23} & s^0_{33} \\
                     \end{array}
                   \right)\sim (6^*,2/3,\underline{3},-4/3),
\eea where the parentheses denote the quantum numbers based on
$(SU(3)_L, U(1)_X, A_4, U(1)_\mathcal{L})$ symmetries,
respectively. The subscripts to the component fields are indices
of $SU(3)_L$. The \underline{3} indices of $A_4$ for $\phi$ and
$s$ are discarded and understood. For convenience, we also list
the lepton number ($L$) for the component particles:

\bc
\begin{tabular}{|c|c|}
  \hline
  Scalars & $L$   \\ \hline
  $\phi^+_1$, $\phi^0_2$, $\eta^0_1$, $\eta^-_2$, $\rho^+_1$, $\rho^0_2$,
  $\chi^0_3$, $\sigma^0_{33}$, $s^0_{33}$ & 0  \\ \hline
   $\phi^+_3$, $\eta^0_3$, $\rho^+_3$, $\chi^{0*}_1$, $\chi^+_2$, $\sigma^0_{13}$,
   $\sigma^+_{23}$, $s^0_{13}$, $s^+_{23}$ & $-1$
   \\ \hline
   $\sigma^{0}_{11}$, $\sigma^{+}_{12}$, $\sigma^{++}_{22}$,
   $s^{0}_{11}$, $s^{+}_{12}$, $s^{++}_{22}$ & $-2$ \\ \hline
\end{tabular}\ec

\subsection{Scalar potential}

We can separate the general scalar potential into \be
V_{\mathrm{scalar}}=V_1+V_2+\bar{V}_3,\ee in which the first and
second term conserves the $\mathcal{L}$ charge whereas the third
term violates this charge.  Moreover, $V_1$ consists of all terms
of $\phi$, $\eta$, $\rho$, $\chi$, without $\sigma$ and $s$; $V_2$
is all the terms having at least a $\sigma$ or $s$. $V_1$ is a sum
of \bea V(\phi)&=&\mu^2_\phi (\phi^\dagger
\phi)_{\underline{1}}+\la^\phi_1 (\phi^\dagger
\phi)_{\underline{1}}(\phi^\dagger \phi)_{\underline{1}}
+\la^\phi_2 (\phi^\dagger \phi)_{\underline{1}'}(\phi^\dagger
\phi)_{\underline{1}''}\crn
 &&+\la^\phi_3 (\phi^\dagger \phi)_{\underline{3_s}}(\phi^\dagger \phi)_{\underline{3_a}}
 +[\la^\phi_4 (\phi^\dagger \phi)_{\underline{3_s}}(\phi^\dagger \phi)_{\underline{3_s}}+h.c.],\\
V(\eta)&=&\mu^2_\eta \eta^\dagger \eta+\la^\eta (\eta^\dagger
\eta)^2,\\
V(\rho)&=&\mu^2_\rho \rho^\dagger \rho+\la^\rho (\rho^\dagger
\rho)^2,\\
V(\chi)&=&\mu^2_\chi \chi^\dagger \chi+\la^\chi (\chi^\dagger
\chi)^2,\\
V(\phi,\eta)&=&\la^{\phi\eta}_1(\phi^\dagger
\phi)_{\underline{1}}(\eta^\dagger \eta)
+\la^{\phi\eta}_2(\phi^\dagger \eta)(\eta^\dagger \phi),\\
V(\phi,\rho)&=&\la^{\phi\rho}_1(\phi^\dagger
\phi)_{\underline{1}}(\rho^\dagger \rho)
+\la^{\phi\rho}_2(\phi^\dagger \rho)(\rho^\dagger \phi)+
[\la^{\phi\rho}_3(\phi^\dagger \rho)(\phi^\dagger \rho)\crn
&&+\la^{\phi\rho}_4(\rho^\dagger \phi)(\phi^\dagger
\phi)_{\underline{3_s}} +\la^{\phi\rho}_5(\rho^\dagger
\phi)(\phi^\dagger
\phi)_{\underline{3_a}}+h.c.],\\
V(\phi,\chi)&=&\la^{\phi\chi}_1(\phi^\dagger
\phi)_{\underline{1}}(\chi^\dagger \chi)
+\la^{\phi\chi}_2(\phi^\dagger \chi)(\chi^\dagger \phi),\\
V(\eta,\rho)&=&\la^{\eta\rho}_1(\eta^\dagger \eta)(\rho^\dagger
\rho)+\la^{\eta\rho}_2(\eta^\dagger \rho)(\rho^\dagger \eta),\\
V(\eta,\chi)&=&\la^{\eta\chi}_1(\eta^\dagger \eta)(\chi^\dagger
\chi)+\la^{\eta\chi}_2(\eta^\dagger \chi)(\chi^\dagger \eta),\\
V(\rho,\chi)&=&\la^{\rho\chi}_1(\rho^\dagger \rho)(\chi^\dagger
\chi)+\la^{\rho\chi}_2(\rho^\dagger \chi)(\chi^\dagger \rho),\\
V(\eta,\rho,\chi)&=&\mu_1 \eta\rho\chi+h.c.\eea

The $V_{2}$ is a sum of \bea
V(s)&=&\mathrm{Tr}\left\{V(\phi\rightarrow s)+\la'^s_1 (s^\dagger
s)_{\underline{1}}\mathrm{Tr}(s^\dagger s)_{\underline{1}}
+\la'^s_2 (s^\dagger s)_{\underline{1}'}\mathrm{Tr}(s^\dagger
s)_{\underline{1}''}\right.\crn &&\left. +\la'^s_3 (s^\dagger
s)_{\underline{3_s}}\mathrm{Tr}(s^\dagger s)_{\underline{3_a}}
+[\la'^s_4 (s^\dagger s)_{\underline{3_s}}\mathrm{Tr}(s^\dagger s)_{\underline{3_s}}+h.c.]\right\},\\
V(\sigma)&=&\mathrm{Tr}[V(\eta\rightarrow \sigma)+\la'^\sigma
(\sigma^\dagger \sigma)\mathrm{Tr}(\sigma^\dagger \sigma)],\\
V(s,\sigma)&=&\mathrm{Tr}\left\{V(\phi\rightarrow s, \rho
\rightarrow \sigma) +\la'^{s\sigma}_1(s^\dagger
s)_{\underline{1}}\mathrm{Tr}(\sigma^\dagger \sigma)
+\la'^{s\sigma}_2(s^\dagger \sigma)\mathrm{Tr}(\sigma^\dagger
s)\right.\crn
&&\left.+[\la'^{s\sigma}_3(s^\dagger\sigma)\mathrm{Tr}(s^\dagger
\sigma)+\la'^{s\sigma}_4(\sigma^\dagger s)\mathrm{Tr}(s^\dagger
s)_{\underline{3_s}} +\la'^{s\sigma}_5(\sigma^\dagger s)(s^\dagger
s)_{\underline{3_a}}\right.\crn &&\left.+h.c.]\right\},\\
V(s,\phi)&=&\mathrm{Tr}\left\{\la^{\phi s}_1(\phi^\dagger
\phi)_{\underline{1}}(s^\dagger s)_{\underline{1}}+\left[\la^{\phi
s}_2(\phi^\dagger \phi)_{\underline{1}'}(s^\dagger
s)_{\underline{1}''}\right.\right.\crn &&\left.+\la^{\phi
s}_3(\phi^\dagger \phi)_{\underline{3_s}}(s^\dagger
s)_{\underline{3_a}}+\la^{\phi s}_4(\phi^\dagger
\phi)_{\underline{3_s}}(s^\dagger
s)_{\underline{3_s}}+h.c.\right]\crn &&+\la^{\phi
s}_5(\phi^\dagger s^\dagger)_{\underline{1}}(s
\phi)_{\underline{1}}+ \la^{\phi s}_6(\phi^\dagger
s^\dagger)_{\underline{1}'}(s \phi)_{\underline{1}''}\crn
&&\left.+\la^{\phi s}_7(\phi^\dagger
s^\dagger)_{\underline{3_s}}(s \phi)_{\underline{3_a}}
+\left[\la^{\phi s}_8(\phi^\dagger s^\dagger)_{\underline{3_s}}
(s \phi)_{\underline{3_s}}+h.c.\right]\right\},\label{vsp}\\
V(s,\eta)&=& \mathrm{Tr}[V(\phi\rightarrow s^\dagger, \eta
\rightarrow \eta)],\\
V(s,\rho)&=& \mathrm{Tr}[V(\phi\rightarrow s^\dagger, \eta
\rightarrow \rho)],\\
V(s,\chi)&=& \mathrm{Tr}[V(\phi\rightarrow s^\dagger, \eta
\rightarrow \chi)],\\
V(\sigma,\phi)&=& \mathrm{Tr}[V(\phi\rightarrow \phi, \eta \rightarrow \sigma^\dagger)],\\
V(\sigma,\eta)&=& \mathrm{Tr}[V(\eta\rightarrow \eta, \rho \rightarrow \sigma^\dagger)],\\
V(\sigma,\rho)&=& \mathrm{Tr}[V(\eta\rightarrow \rho, \rho
\rightarrow \sigma^\dagger)],\\
V(\sigma,\chi)&=&\mathrm{Tr}[V(\eta\rightarrow \chi, \rho
\rightarrow \sigma^\dagger)]+[\mu_2 \chi^T \sigma \chi+h.c.],\\
V(s,\phi,\eta,\chi)&=& \la_{1}\chi^\dagger s^\dagger\eta\phi +h.c.
 \eea Notice that
 $(\mathrm{Tr}A)(\mathrm{Tr}B)=\mathrm{Tr}(A\mathrm{Tr}B)$, and $V(X\rightarrow X_1,Y\rightarrow
 Y_1)\equiv V(X,Y)|_{X=X_1,Y=Y_1}$.

The third term $\bar{V}_3$ is given by \bea
\bar{V}_3&=&\bar{\mu}_{1} \eta^T \sigma \chi+\bar{\mu}_{2} \eta^T
 \sigma\eta+\bar{\la}_{1} \eta^\dagger
 s^\dagger\chi\phi+\bar{\la}_{2} \eta^\dagger
 s^\dagger\eta\phi+\bar{\la}_{3} \chi^\dagger
 s^\dagger\chi\phi + \bar{\la}_{4} \eta^\dagger s^\dagger s \chi
 + \bar{\la}_{5} \eta^\dagger \sigma^\dagger\sigma \chi\crn &&+\left[\bar{\la}_{6}
\mathrm{Tr}(\sigma^\dagger \sigma) +\bar{\la}_{7}
 \mathrm{Tr}(s^\dagger s)+ \bar{\la}_8
\eta^\dagger\chi + \bar{\la}_9 \eta^\dagger\eta +\bar{\la}_{10}
\rho^\dagger\rho+\bar{\la}_{11} \chi^\dagger\chi+\bar{\la}_{12}
\phi^\dagger\phi +\bar{\mu}^2_3 \right] \eta^\dagger \chi \crn &&+ \bar{\la}_{13}
 (\eta^\dagger \rho)(\rho^\dagger \chi)
 + \bar{\la}_{14} (\eta^\dagger \phi)(\phi^\dagger \chi)+h.c.\label{vi}
\eea There may exist soft -terms in $\bar{V}$ explicitly violating
the $A_4$ symmetry. But, only some of them are  mentioned in the
text.

\end{document}